\documentclass[12pt]{article}
\usepackage{graphicx}
\begin{document}

\begin{center}
{\Large \bf Coupled oscillators, entangled oscillators, and Lorentz-covariant
harmonic oscillators}

\vspace{3ex}

Y. S. Kim\footnote{electronic address: yskim@physics.umd.edu}\\
Department of Physics, University of Maryland,\\
College Park, Maryland 20742, U.S.A.\\

\vspace{3ex}

Marilyn E. Noz \footnote{electronic address: noz@nucmed.med.nyu.edu}\\
Department of Radiology, New York University,\\ New York, New York 10016, U.S.A.\\

\end{center}

\vspace{3ex}

\vspace{20mm}

\begin{abstract}

Other than scattering problems where perturbation theory is applicable,
there are basically two ways to solve problems in physics.  One is
to reduce the problem to harmonic oscillators, and the other is to
formulate the problem in terms of two-by-two matrices.  If two
oscillators are coupled, the problem combines both two-by-two matrices
and harmonic oscillators.  This method then becomes a powerful research
tool to cover many different branches of physics.
Indeed, the concept and methodology in one branch of physics can
be translated into another through the common mathematical formalism.
Coupled oscillators provide clear illustrative examples for some
of the current issues in physics, including entanglement and Feynman's
rest of the universe.
In addition, it is noted that the present form of quantum mechanics
is largely a physics of harmonic oscillators.  Special relativity is
the physics of the Lorentz group which can be represented by the group
of two-by-two matrices commonly called $SL(2,c)$.  Thus the coupled
harmonic oscillators can play the role of combining quantum mechanics
with special relativity.
It is therefore possible to relate the current issues of physics to
the Lorentz-covariant formulation of quantum mechanics.
\end{abstract}

\vspace{10mm}
PACS: 03.65.Ud, 03.65.Yz, 11.10.St, 11.30.Cp

\newpage

\section{Introduction}\label{intro}

Because of its mathematical simplicity, the harmonic oscillator provides
soluble models in many branches of physics.  It often gives a clear
illustration of abstract ideas.  In many cases, the problems are reduced
to the problem of two coupled oscillators.  Soluble models in quantum
field theory, such as the Lee model~\cite{sss61} and the Bogoliubov
transformation in superconductivity~\cite{fewa71}, are based on two
coupled oscillators.    Recently, two coupled oscillators formed the
mathematical basis for squeezed states in quantum optics~\cite{knp91},
especially two-mode squeezed states\cite{dir63,vourdas88}.

More recently, it was noted by Giedke {\it et al.} that entanglement
realized in two-mode squeezed states can be formulated in terms of
symmetric Gaussian states~\cite{giedke03}.
From a mathematical point of view, the subject of entanglement has
been largely a physics of two-by-two matrices.  It is gratifying to
note that harmonic oscillators can also play a role in clarifying
the physical basis of entanglement.
The symmetric Gaussian states can be constructed from two coupled
harmonic oscillators.  The entanglement issues in two-mode squeezed
states can therefore be added to the physics of coupled harmonic
oscillators.  Since many physical models are based on coupled
oscillators, entanglement ideas can be exported to all those models.

In this paper, we construct a model of Lorentz-covariant harmonic
oscillators based on the coupled oscillators.  Since the covariance
requires coupling of space and time variables, the covariant
oscillator formalism allows expansion of entanglement ideas to
the space-time region.

Combining quantum mechanics with special relativity is a fundamental
problem in its own right.  Why do we need covariant harmonic oscillators
while there is quantum field theory with Feynman diagrams?  Since this
is also a fundamental problem by its own right, we would like to address
this issue in the first sections of this paper.  The point is that
quantum mechanics deals with waves.  There are running waves and
standing waves.  The present form of quantum mechanics and its S-matrix
formalism are only for running waves, and cannot directly deal with
standing waves satisfying boundary conditions.  Of course standing waves
are superpositions of two running waves, but we do not know how to
approach this problem when we include Lorentz transformations.  The
simplest approach is to work with a soluble model based on harmonic
oscillators and two-by-two matrices.

From the mathematical  point of view, special relativity is a physics
of Lorentz transformations or the Lorentz group.  It is gratifying to
note that the six-parameter Lorentz group can be represented by
two-by-two matrices with unit determinant.  The elements can be complex
numbers.  This group is known as $SL(2,C)$ which forms the universal
covering group of the Lorentz group.  Thus, special relativity is a
physics of two-by-two matrices.

The standard approach to two coupled oscillators is to construct a
two-by-two matrix of two oscillators with different frequencies.  Thus,
it is not surprising to note that the mathematics of the coupled
oscillators is directly applicable to Lorentz-covariant harmonic
oscillators.  Therefore, the covariant oscillators, defined in the
space-time region, can be enriched by the physics of entanglement.

As for using coupled harmonic oscillators for combining quantum
mechanics with relativity, we examine in this paper the earlier attempts
made by Dirac and Feynman.  We first examine Dirac's approach which was
to construct mathematically appealing models.  We then examine how Feynman
approached this problem.  He observed the experimental world, told the
story of the real world in his style, and then wrote down mathematical
formulas as needed.  We use coupled oscillators to combine Dirac's
approach and Feynman's approach to construct the Lorentz-covariant
formulation of quantum mechanics.

In Sec.~\ref{scabo}, it is noted that quantum mechanics deals with waves,
and there are running waves and standing waves.  While it is somewhat
straightforward to make running waves Lorentz-covariant, there are no
established prescriptions for constructing standing waves consistent
with special relativity.  We stress in this section, why standing waves
are different from running waves.  In Sec.~\ref{quantu}, we discuss
the quantum mechanics of two coupled oscillators, and study how the
system could absorb the physical ideas developed in the case of two-mode
squeezed states.

In Sec.~\ref{dirosc}, we study systematically Dirac's lifetime efforts
to combine quantum mechanics with relativity.  He was concerned with
space-time asymmetry associated with position-momentum and time-energy
uncertainty relations.  We examine carefully what more had to be done
to complete the task initiated by Dirac.  In Sec.~\ref{feyosc}, we study
Feynman's efforts to combine quantum mechanics with special relativity.
Here also, we carefully examine the short-comings in Feynman's papers
on harmonic oscillators.  It is shown possible in Sec.~\ref{covham} that
the works of Feynman and Dirac can be combined to produce a covariant
harmonic oscillator system.  In Sec.~\ref{sten}, it is shown that the
covariant oscillator formalism shares the same mathematical base as that
of two coupled oscillators, and much of the physical ideas, especially
the entanglement idea, can be translated into the space-time variables
of the Lorentz-covariant world.

The physics of space-time has its own merit, and is not bound to import
ideas developed in other areas of physics.  In Sec.~\ref{feydeco}, we
note that there is a dechorence effect observed first by Feynman.  It
is known widely as Feynman's parton picture in which partons appear
like incoherent entities.  It is widely believed that partons are
Lornents-boosted quarks.  Then, the question is how the Lorentz boost,
which is a space-time symmetry operation, can destroy coherence.  We
address this question in this section,

\section{Scattering States and Bound States}\label{scabo}

In this section, we would like to address the question of why we need
covariant harmonic oscillators while there is the Lorentz-covariant
formulation of quantum field theory which allows us to calculate
scattering amplitudes using Feynman diagrams.

When Einstein formulated his special relativity one hundred years ago,
he was considering point particles. Einstein's energy-momentum relation
is known to be valid also for particles with space-time extensions.
There have been efforts to understand special relativity for rigid
particles with non-zero size, without any tangible result.  On the
other hand, the emergence of quantum mechanics made the rigid-body
problem largely irrelevant.
Thanks to wave-particle duality, we talk about wave packets and
standing waves, instead of rigid bodies.  The issue
becomes whether those waves can be made Lorentz-covariant.

Of course, here, the starting point is the plane wave,
which can be written as
\begin{equation}
e^{ip \cdot x} = e^{i(\vec{p}\cdot\vec{x} - Et)} .
\end{equation}
Since it takes the same form for all Lorentz frames, we do not
need any extra effort to make it Lorentz covariant.

Indeed, the S-matrix derivable from the present form of quantum
field theory calls for calculation of all S-matrix quantities
in terms of plane waves. Thus, the S-matrix is associated with
perturbation theory or Feynman diagrams.  Feynman
propagators are written in terms of  plane waves on the mass
shell.

\begin{figure}[thb]
\centerline{\includegraphics[scale=0.7]{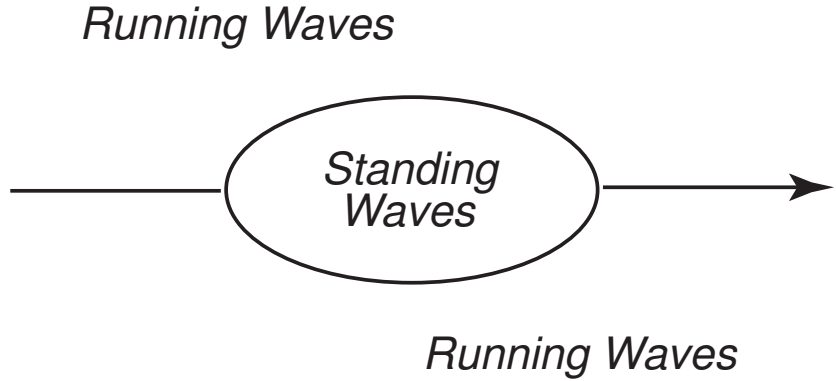}}
\vspace{5mm}
\caption{Running waves and standing waves in quantum theory.  If a
particle is allowed to travel from infinity to infinity, it corresponds
to a running wave according to the wave picture of quantum mechanics.
If, on the other hand, it is trapped in a localized region, we have
to use standing waves to interpret its location in terms of
probability distribution.}\label{dff11}
\end{figure}

\begin{figure}[thb]
\centerline{\includegraphics[scale=0.5]{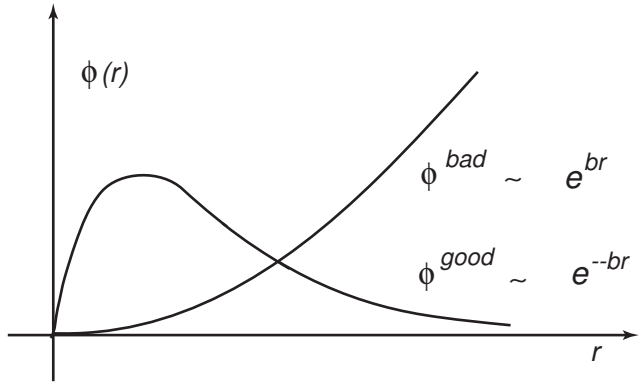}}
\vspace{5mm}
\caption{Good and bad wave functions contained in the S-matrix.
Bound-state wave functions satisfy the localization condition and are
good wave functions.  Analytic continuations of plane waves do not
satisfy the localization boundary condition, and become bad wave
functions at the bound-state
energy.}\label{goodbad}
\end{figure}

We should realize however that the S-matrix formalism is strictly
for running waves, starting from a plane wave from one end of the
universe and ending with another plane wave at another end.  How
about standing waves?  This question is illustrated in
Fig.~\ref{dff11}.  Of course, standing waves can be regarded as
superpositions of running waves moving in opposite directions.
However, in order to guarantee localization of the standing waves,
we need a spectral function or boundary conditions.  The covariance
of standing waves necessarily involves the covariance of boundary
conditions or spectral functions.  How much do we know about this
problem?  This problem has not yet been systematically explored.

On the other hand, some concrete models for covariant bound-states
were studied in the past by a number of distinguished physicists,
including Paul A. M. Dirac~\cite{dir45}, Hideki Yukawa~\cite{yuka53},
and Richard Feynman and his co-authors~\cite{fkr71}.  We shall return
to this problem in Sec.~\ref{dirosc}.

Finally, let us see what kind of problems we expect if we use S-matrix
methods for bound state problems.  If we use the spherical coordinate
system where the scattering center is at the origin, the S-matrix
consists of both incoming and outgoing waves.  If we make analytic
continuations of these waves to bound states with negative total energy,
the outgoing wave becomes localized, but the incoming wave increases
to infinity at large distance from the origin, as indicated in
Fig.~\ref{goodbad}.  There no methods of eliminating this unphysical
wave function.

Indeed, this was the source of mistake made by Dashen and Frautschi in
their once-celebrated calculation of the neutron and proton mass
difference using an S-matrix formula corresponding to the first-order
energy shift~\cite{kim66}.  They used a pertubation formula derivable
from S-matrix considerations, but their formula corresponds to the
pertubation formula:
\begin{equation}\label{shift}
\delta E = \left(\phi^{good}, \delta V \phi^{bad} \right) ,
\end{equation}
where the good and bad bound-state wave functions are like
\begin{equation}\label{wfs}
 \phi^{good} \sim e^{-br} , \qquad
\phi^{bad} \sim e^{br} ,
\end{equation}
for large values of $r$, as illustrated in Fig.~\ref{goodbad}.  We are
not aware of any S-matrix method which gurantees the localization of
bound-state wave functions.

\section{Coupled Oscillators and Entangled Oscillators}\label{quantu}

The coupled oscillator problem can be formulated as that of a
quadratic equation in two variables.  The diagonalization
of the quadratic form includes a rotation of the coordinate system.
However, the diagonalization process requires
additional transformations involving the scales of the coordinate
variables~\cite{arav89,hkn99ajp}.  Indeed, it was found that the
mathematics of this procedure can be as complicated as the group
theory of Lorentz transformations in a six dimensional space with
three spatial and three time coordinates~\cite{hkn95jm}.

In this paper, we start with a simple problem of two identical
oscillators. Then the Hamiltonian takes the form
\begin{equation}
H = {1\over 2}\left\{{1\over m} p^{2}_{1} + {1\over m}p^{2}_{2}
+ A x^{2}_{1} + A x^{2}_{2} + 2C x_{1} x_{2} \right\}.
\end{equation}
If we choose coordinate variables
\begin{eqnarray}\label{normal}
&{}& y_{1} = {1\over\sqrt{2}}\left(x_{1} + x_{2}\right) , \nonumber\\[2ex]
&{}& y_{2} = {1\over\sqrt{2}}\left(x_{1} - x_{2}\right) ,
\end{eqnarray}
the Hamiltonian can be written as
\begin{equation}\label{eq.6}
H = {1\over 2m} \left\{p^{2}_{1} + p^{2}_{2} \right\} +
{K\over 2}\left\{e^{-2\eta} y^{2}_{1} + e^{2\eta} y^{2}_{2} \right\} ,
\end{equation}
where
\begin{eqnarray}
&{}&   K = \sqrt{A^{2} - C^{2}} ,  \nonumber \\[.5ex]
&{}& \exp(2\eta) =\sqrt{\frac{A - C}{A + C} } ,
\end{eqnarray}
The classical eigenfrequencies are $\omega_{\pm} = \omega e^{{\pm}2\eta}$
with $\omega = \sqrt{K/m}$ .

If $y_{1}$ and $y_{2}$ are measured in units of $(mK)^{1/4} $,
the ground-state wave function of this oscillator system is
\begin{equation}\label{wf01}
\psi_{\eta}(x_{1},x_{2}) = {1 \over \sqrt{\pi}}
\exp{\left\{-{1\over 2}(e^{-2\eta} y^{2}_{1} + e^{2\eta} y^{2}_{2})
\right\} } ,
\end{equation}
The wave function is separable in the $y_{1}$ and $y_{2}$ variables.
However, for the variables $x_{1}$ and $x_{2}$, the story is quite
different, and can be extended to the issue of entanglement.

There are three ways to excite this ground-state oscillator system.
One way is to multiply Hermite polynomials for the usual quantum
excitations.  The second way is to construct coherent states for
each of the $y$ variables.  Yet, another way is to construct
thermal excitations.  This requires density matrices and Wigner
functions~\cite{hkn99ajp}.

The key question is how the quantum mechanics in the world of the
$x_{1}$ variable is affected by the $x_{2}$ variable.
If we use two separate measurement processes for
these two variables, these two oscillators are entangled.

Let us write the wave function of Eq.(\ref{wf01}) in terms of
$x_{1}$ and $x_{2}$, then
\begin{equation}\label{wf02}
\psi_{\eta}(x_{1},x_{2}) = {1 \over \sqrt{\pi}}
\exp\left\{-{1\over 4}\left[e^{-2\eta}(x_{1} + x_{2})^{2} +
e^{2\eta}(x_{1} - x_{2})^{2} \right] \right\} .
\end{equation}
When the system is decoupled with $\eta = 0$, this wave function becomes
\begin{equation}
\psi_{0}(x_{1},x_{2}) = \frac{1}{\sqrt{\pi}}
\exp{\left\{-{1\over 2}(x^{2}_{1} + x^{2}_{2}) \right\}} .
\end{equation}
The system becomes separable and becomes disentangled.

As was discussed in the literature for several different
purposes~\cite{knp91,kno79ajp,knp86}, this wave function can be
expanded as
\begin{equation}\label{expan1}
\psi_{\eta }(x_{1},x_{2}) = {1 \over \cosh\eta}\sum^{}_{k}
(\tanh\eta )^{k} \phi_{k}(x_{1}) \phi_{k}(x_{2}) ,
\end{equation}
where $\phi_{k}(x)$ is the normalized harmonic oscillator wave function for the
$k-th$ excited state.
This expansion serves as the mathematical basis for squeezed states
of light in quantum optics~\cite{knp91}, among other applications.

The expansion given in Eq.(\ref{expan1}) clearly demonstrates that
the coupled oscillators are entangled oscillators.  This expression
is identical to Eq.(1) of the recent paper by Giedke
{\it et al.}~\cite{giedke03}.  This means that the coupled oscillators
can absorb most of the current entanglement issues, and serve as a
reservior of entanglment ideas for other physical systems modeled
after the coupled oscillators.  We are particularly interested in
expanding these ideas to relativistic space and time through the
covariant oscillator formalism.

In Sec~\ref{dirosc}, we shall see that the mathematics of the coupled
oscillators can serve as the basis for the covariant harmonic
oscillator formalism where the $x_{1}$ and $x_{2}$ variables
are replaced by the longitudinal and time-like variables,
respectively.  This mathematical identity will lead to the concept
of space-time entanglement in special relativity, as we shall
see in Sec.~\ref{sten}.

\section{Dirac's Harmonic Oscillators and Light-cone Coordinate
System}\label{dirosc}

Paul A. M. Dirac is known to us through the Dirac equation for spin-1/2
particles.  But his main interest was in the foundational problems.
First, Dirac was never satisfied with the probabilistic formulation of
quantum mechanics.  This is still one of the hotly debated subjects in
physics.  Second, if we tentatively accept the present form of quantum
mechanics, Dirac was insisting that it has to be consistent with
special relativity.  He wrote several important papers on this subject.
Let us look at some of his papers.

\begin{figure}[thb]
\centerline{\includegraphics[scale=0.8]{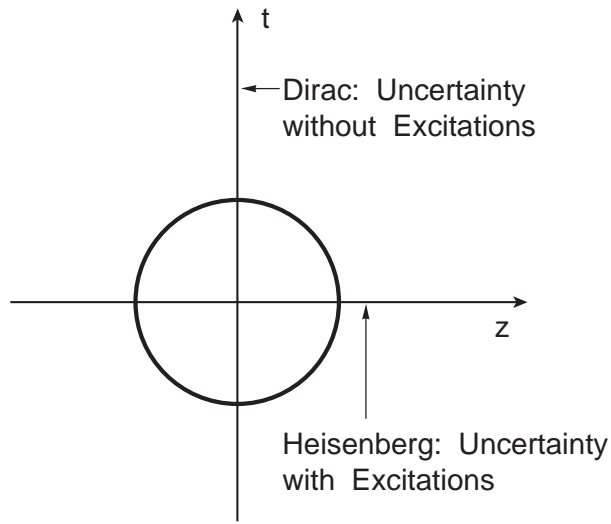}}
\vspace{5mm}
\caption{Space-time picture of quantum mechanics.  There
are quantum excitations along the space-like longitudinal direction, but
there are no excitations along the time-like direction.  The time-energy
relation is a c-number uncertainty relation.}\label{quantum}
\end{figure}

During World War II, Dirac was looking into the possibility of constructing
representations of the Lorentz group using harmonic oscillator wave
functions~\cite{dir45}.  The Lorentz group is the language of special
relativity, and the present form of quantum mechanics starts with harmonic
oscillators.  Presumably, therefore, he was interested in making quantum
mechanics Lorentz-covariant by constructing representations of the Lorentz
group using harmonic oscillators.

In his 1945 paper~\cite{dir45}, Dirac considered the Gaussian form
\begin{equation}
\exp\left\{- {1 \over 2}\left(x^2 + y^2 + z^2 + t^2\right)\right\} .
\end{equation}
This Gaussian form is in the $(x,~y,~z,~t)$
coordinate variables.  Thus, if we consider Lorentz boost along the
$z$ direction, we can drop the $x$ and $y$ variables, and write the
above equation as
\begin{equation}\label{ground}
\exp\left\{- {1 \over 2}\left(z^2 + t^2\right)\right\} .
\end{equation}
This is a strange expression for those who believe in Lorentz invariance.
The expression $\left(z^2 + t^2\right)$ is not invariant under Lorentz
boost.  Therefore  Dirac's Gaussian form of Eq.(\ref{ground}) is not
a Lorentz-invariant expression.

On the other hand, this expression is consistent with his earlier papers
on the time-energy uncertainty relation~\cite{dir27}.  In those papers,
Dirac observed that there is a time-energy uncertainty relation, while
there are no excitations along the time axis.  He called this the
``c-number time-energy uncertainty'' relation.  When one of us (YSK) was
talking with Dirac in 1978, he clearly mentioned this word again.  He
said further that this space-time asymmetry is one of the stumbling block
in combining quantum mechanics with relativity. This situation is
illustrated in Fig.~\ref{quantum}.

\begin{figure}[thb]
\centerline{\includegraphics[scale=0.8]{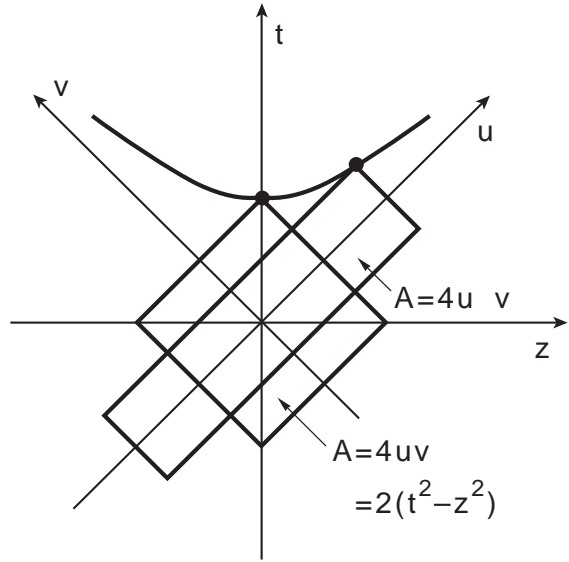}}
\vspace{5mm}
\caption{Lorentz boost in the light-cone coordinate
system.}\label{licone}
\end{figure}

In 1949, the Reviews of Modern Physics published a special issue to
celebrate Einstein's 70th birthday.  This issue contains Dirac paper
entitled ``Forms of Relativistic Dynamics''~\cite{dir49}.
In this paper, he introduced his light-cone coordinate system,
in which a Lorentz boost becomes a squeeze transformation.

When the system is boosted along the $z$ direction, the transformation
takes the form
\begin{equation}\label{boostm}
\pmatrix{z' \cr t'} = \pmatrix{\cosh\eta & \sinh\eta \cr
\sinh\eta & \cosh\eta } \pmatrix{z \cr t} .
\end{equation}
The light-cone variables are defined as~\cite{dir49}
\begin{equation}\label{lcvari}
u = (z + t)/\sqrt{2} , \qquad v = (z - t)/\sqrt{2} ,
\end{equation}
the boost transformation of Eq.(\ref{boostm}) takes the form
\begin{equation}\label{lorensq}
u' = e^{\eta } u , \qquad v' = e^{-\eta } v .
\end{equation}
The $u$ variable becomes expanded while the $v$ variable becomes
contracted, as is illustrated in Fig.~\ref{licone}.  Their product
\begin{equation}
uv = {1 \over 2}(z + t)(z - t) = {1 \over 2}\left(z^2 - t^2\right)
\end{equation}
remains invariant.  In Dirac's picture, the Lorentz boost is a
squeeze transformation.

\begin{figure}[thb]
\centerline{\includegraphics[scale=0.4]{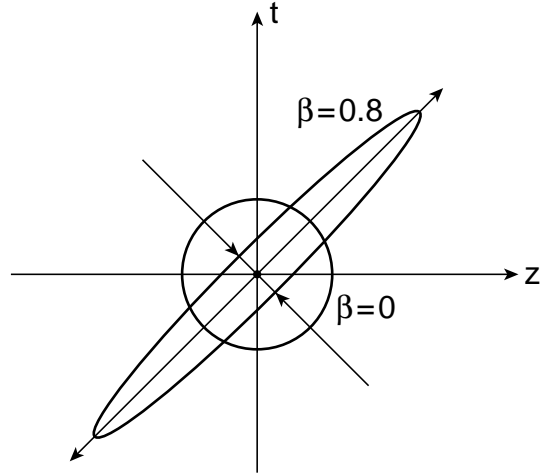}}
\caption{Effect of the Lorentz boost on the space-time
wave function.  The circular space-time distribution in the rest frame
becomes Lorentz-squeezed to become an elliptic
distribution.}\label{ellipse}
\end{figure}

If we combine Fig.~\ref{quantum} and Fig.~\ref{licone}, then we end up
with Fig.~\ref{ellipse}.
This transformation changes the Gaussian form
of Eq.(\ref{ground}) into
\begin{equation}\label{eta}
\psi_{\eta }(z,t) = \left({1 \over \pi }\right)^{1/2}
\exp\left\{-{1\over 2}\left(e^{-2\eta }u^{2} +
e^{2\eta}v^{2}\right)\right\} .
\end{equation}
Let us go back to Sec.~\ref{quantu} on the coupled oscillators.  The
above expression is the same as Eq.(\ref{wf01}).  The $x_{1}$ variable
now became the longitudinal variable $z$, and the $x_{2}$ variable
became the time like variable $t$.

We can use the coupled harmonic oscillators as the starting point of
relativistic quantum mechanics.  This allows us to translate the quantum
mechanics of two coupled oscillators defined over the space of $x_{1}$
and $x_{2}$ into quantum mechanics defined over the space-time
region of $z$ and $t$.

This form becomes (\ref{ground}) when $\eta$ becomes zero.  The
transition from Eq.(\ref{ground}) to Eq.(\ref{eta}) is a squeeze
transformation.  It is now possible to combine what Dirac observed
into a covariant formulation of the harmonic oscillator system. First,
we can combine his c-number time-energy uncertainty relation described
in Fig.~\ref{quantum} and his light-cone coordinate system of
Fig.~\ref{licone} into a picture of covariant space-time localization
given in Fig.~\ref{ellipse}.

The wave function of Eq.(\ref{ground}) is distributed within a
circular region in the $u v$ plane, and thus in the $z t$ plane.
On the other hand, the wave function of Eq.(\ref{eta}) is distributed
in an elliptic region with the light-cone axes as the major and
minor axes respectively.  If $\eta$ becomes very large, the wave
function becomes concentrated along one of the light-cone axes.
Indeed, the form given in Eq.(\ref{eta}) is a Lorentz-squeezed wave
function.  This squeeze mechanism is illustrated in Fig.~\ref{ellipse}.

There are two homework problems which Dirac left us to solve. First,
in defining the $t$ variable for the Gaussian form of Eq.(\ref{ground}),
Dirac did not specify the physics of this variable.
If it is going to be the calendar time, this form vanishes in the remote
past and remote future.  We are not dealing with this kind of object in
physics.  What is then the physics of this time-like $t$ variable?

The Schr\"odinger quantum mechanics of the hydrogen atom deals with
localized probability distribution.  Indeed, the localization condition
leads to the discrete energy spectrum.  Here, the uncertainty relation
is stated in terms of the spatial separation between the proton and
the electron.  If we believe in Lorentz covariance, there must also
be a time-separation between the two constituent particles, and an
uncertainty relation applicable to this separation variable.  Dirac
did not say in his papers of 1927 and 1945, but Dirac's ``t'' variable
is applicable to this time-separation variable.  This time-separation
variable will be discussed in detail in Sec.~\ref{feyosc} for the
case of relativistic extended particles.

Second, as for the time-energy uncertainty relation,  Dirac'c
concern was how the c-number time-energy uncertainty relation without
excitations can be combined with uncertainties in the position space
with excitations.  How can this space-time asymmetry be consistent
with the space-time symmetry of special relativity?

Both of these questions can be answered in terms of the space-time
symmetry of bound states in the Lorentz-covariant regime~\cite{knp86}.
In his 1939 paper~\cite{wig39}, Wigner worked out internal space-time
symmetries of relativistic particles.  He approached the problem by
constructing the maximal subgroup of the Lorentz group whose
transformations leave the given four-momentum invariant.  As a
consequence, the internal symmetry of a massive particle is like the
three-dimensional rotation group which does not require transformation
into time-like space.

If we extend Wigner's concept to relativistic bound states, the
space-time asymmetry which Dirac observed in 1927 is quite consistent
with Einstein's Lorentz covariance~\cite{hnks83}. Indeed, Dirac's time
variable can be treated separately.  Furthermore, it is possible to
construct a representations of Wigner's little group for massive
particles~\cite{knp86}.  As for the time-separation which can be
linearly mixed with space-separation variables when the system is
Lorentz-boosted, it has its role in internal space-time symmetry.

Dirac's interest in harmonic oscillators did not stop with his 1945
paper on the representations of the Lorentz group.  In his
1963 paper~\cite{dir63}, he constructed a representation of the
$O(3,2)$ deSitter group using two coupled harmonic oscillators.
This paper contains not only the mathematics of combining special
relativity with the quantum mechanics of quarks inside hadrons, but
also forms the foundations of two-mode squeezed states which are so
essential to modern quantum optics~\cite{knp91,vourdas88}.   Dirac
did not know this when he was writing this 1963 paper.  Furthermore,
the $O(3,2)$ deSitter group contains the Lorentz group $O(3,1)$ as a
subgroup.  Thus, Dirac's oscillator representation of the deSitter
group essentially contains all the mathematical ingredient of what
we are studying in this paper.

\section{Feynman's Oscillators }\label{feyosc}
\begin{figure}[thb]
\centerline{\includegraphics[scale=0.7]{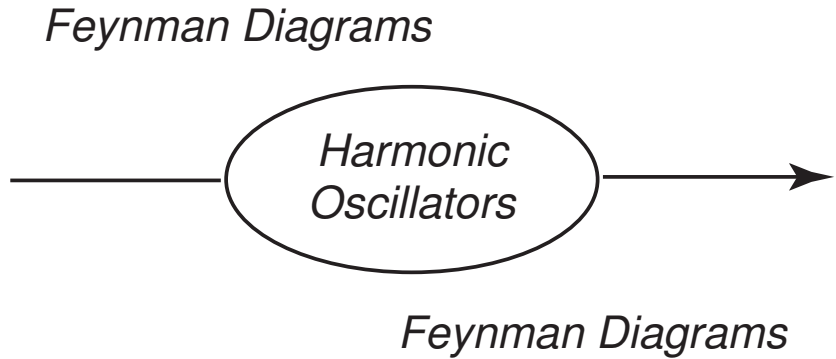}}
\vspace{5mm}
\caption{Feynman's roadmap for combining quantum mechanics with special
relativity.  Feynman diagrams work for running waves, and they provide
a satisfactory resolution for scattering states in Einstein's world.
For standing waves trapped inside an extended hadron, Feynman suggested
harmonic oscillators as the first step.}\label{dff33}
\end{figure}

In his invited talk at the 1970 spring meeting of the American Physical
Society~\cite{feyn70}, Feynman was addressing hadronic mass spectra and
a possible covariant formulation of harmonic oscillators.  He noted that
the mass spectra are consistent with degeneracy of three-dimensional
harmonic oscillators.  Furthermore, Feynman stressed that Feynman
diagrams are not necessarily suitable for relativistic bound states and
that we should try harmonic oscillators.  Feynman's point was that, while
plane-wave approximations in terms of Feynman diagrams work well for
relativistic scattering problems, they are not applicable to bound-state
problems.  We can summarize what Feynman said in Fig.~\ref{goodbad} and
Fig.~\ref{dff33}.

In their 1971 paper~\cite{fkr71}, Feynman, Kislinger and Ravndal started
their harmonic oscillator formalism by defining coordinate variables for
the quarks confined within a hadron.  Let us use the simplest hadron
consisting of two quarks bound together with an attractive force, and
consider their space-time positions $x_{a}$ and $x_{b}$, and use the
variables
\begin{equation}
X = (x_{a} + x_{b})/2 , \qquad x = (x_{a} - x_{b})/2\sqrt{2} .
\end{equation}
The four-vector $X$ specifies where the hadron is located in space and
time, while the variable $x$ measures the space-time separation
between the quarks.  According to Einstein, this space-time separation
contains a time-like component which actively participates as in
Eq.(\ref{boostm}), if the hadron is boosted along the $z$ direction.
This boost can be conveniently described by the light-cone variables
defined in Eq(\ref{lcvari}).

What do Feynman {\it et al.} say about this oscillator wave function?
In their classic 1971 paper~\cite{fkr71}, they start with the following
Lorentz-invariant differential equation.
\begin{equation}\label{osceq}
{1\over 2} \left\{x^{2}_{\mu} -
{\partial^{2} \over \partial x_{\mu }^{2}}
\right\} \psi(x) = \lambda \psi(x) .
\end{equation}

This partial differential equation has many different solutions
depending on the choice of separable variables and boundary conditions.
Feynman {\it et al.} insist on Lorentz-invariant solutions which are
not normalizable.  On the other hand, if we insist on normalization,
the ground-state wave function takes the form of Eq.(\ref{ground}).
It is then possible to construct a representation of the
Poincar\'e group from the solutions of the above differential
equation~\cite{knp86}.  If the system is boosted, the wave function
becomes given in Eq.(\ref{eta}).

Although this paper contained the above mentioned original idea of Feynman,
it contains some serious mathematical flaws.
Feynman {\it et al.} start with a Lorentz-invariant differential equation
for the harmonic oscillator for the quarks bound together inside a hadron.
For the two-quark system, they write the wave function of the form
\begin{equation}
\exp{\left\{-{1 \over 2}\left(z^2 - t^2 \right) \right\}} ,
\end{equation}
where $z$ and $t$ are the longitudinal and time-like separations between the
quarks.  This form is invariant under the boost, but is not normalizable in
the $t$ variable.  We do not know what physical interpretation to give to
this the above expression.

On the other hand, Dirac's Gaussian form given in Eq.(\ref{ground})
also satisfies Feynman's Lorentz-invariant differential equation.
This Gaussian function is normalizable, but is not invariant under
the boost.  However, the word ``invariant'' is quite different
from the word ``covariant.''  The above form can be covariant
under Lorentz transformations.  We shall get back to this
problem in Sec.~\ref{covham}.

Feynman {\it et al.} studied in detail the degeneracy of the
three-dimensional harmonic oscillators, and compared their results
with the observed experimental data.  Their work is complete and
thorough, and is consistent with the $O(3)$-like symmetry dictated
by Wigner's little group for massive particles~\cite{knp86,wig39}.
Yet, Feynman {\it et al.} make an apology that the symmetry is
not $O(3,1)$.  This unnecessary apology causes a confusion not
only to the readers but also to the authors themselves, and makes
the paper difficult to read.

\section{Can harmonic oscillators be made covariant?}\label{covham}

The simplest solution to the differential equation of Eq.(\ref{osceq})
takes the form of Eq.(\ref{ground}).  If we allow excitations along
the longitudinal coordinate and forbid excitations along the time
coordinate, the wave function takes the form
\begin{equation}\label{wf1}
\psi^{n}_{0} (z,t) = C_{n}        H_{n}(z)
\exp{\left\{- {1 \over 2}\left(z^2 + t^2\right)\right\}} ,
\end{equation}
where $H_{n}$ is the Hermite polynomial of the n-th order, and $C_{n}$
is the normalization constant.

If the system is boosted along the z direction, the $z$ and $t$ variables
in the above wave function should be replaced by $z'$ and $t'$ respectively
with
\begin{equation}
z' = (\cosh\eta) z - (\sinh\eta) t , \qquad
t' = (\cosh\eta) t - (\sinh\eta) z .
\end{equation}
The Lorentz-boosted wave function takes the form
\begin{equation}\label{wf2}
\psi^{n}_{\eta} (z,t) = H_{n}(z')
\exp\left\{- {1 \over 2}\left(z'^2 + t'^2\right)\right\} ,
\end{equation}

\begin{figure}[thb]
\centerline{\includegraphics[scale=0.4]{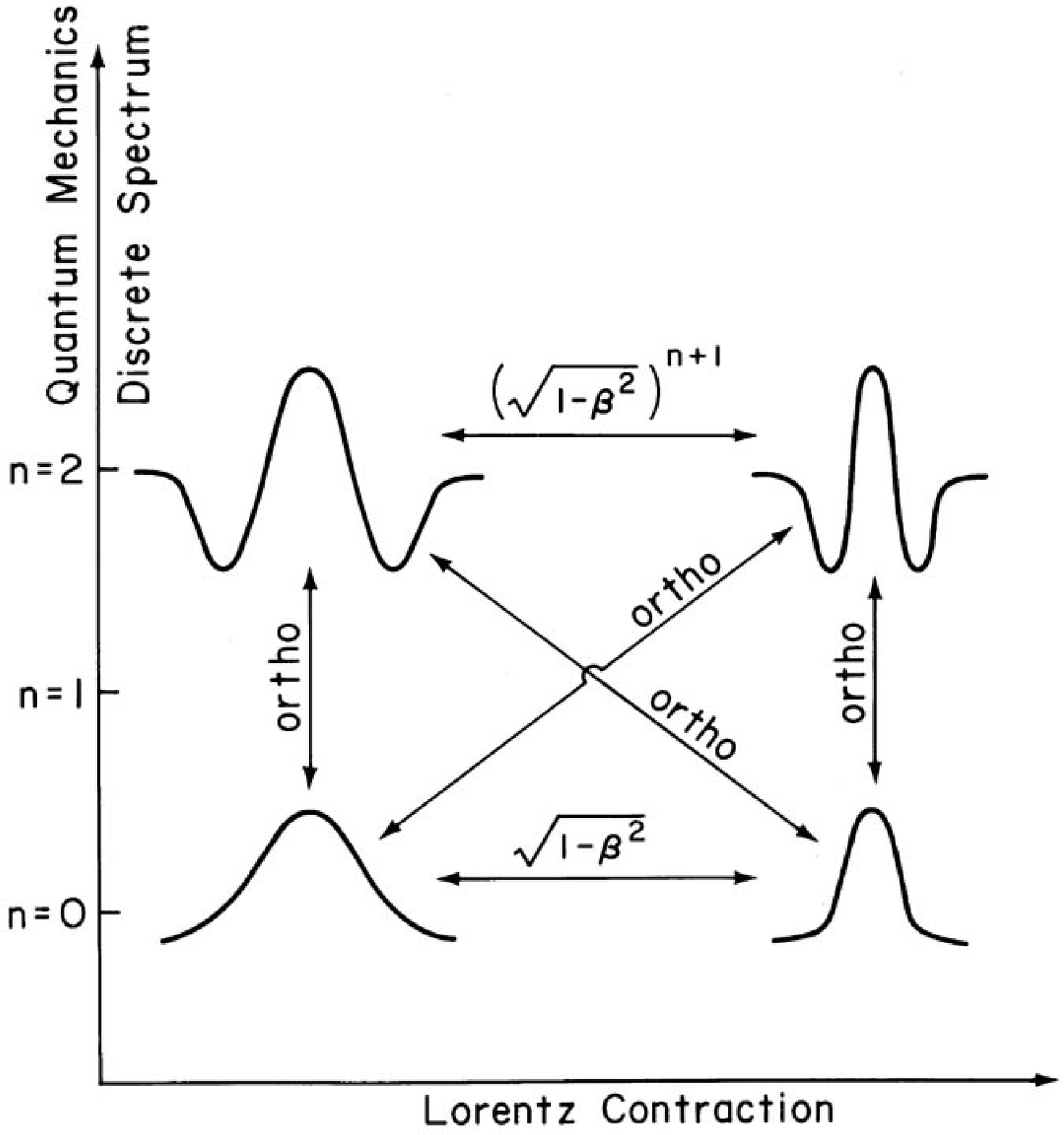}}
\caption{Orthogonality relations for covariant oscillator wave functions.
The orthogonality relations remain invariant under Lorentz boosts, but
their inner products have interesting contraction properties.}\label{ortho}
\end{figure}

It is interesting that these wave functions satisfy the orthogonality
condition~\cite{ruiz74}.
\begin{equation}
\int \psi^{n}_{0} (z,t) \psi^{m}_{\eta} (z,t)dz dt =
\left(\sqrt{1 - \beta^2}\right)^{n} \delta_{nm} ,
\end{equation}
where $\beta = \tanh \eta $.  This orthogonality relation is
illustrated in Fig.~\ref{ortho}.  The physical interpretation of this
in terms of Lorentz contractions is given in our book~\cite{knp86}, but
seems to require further investigation.

It is indeed possible to construct the representation of Wigner's
$O(3)$-like little group for massive particles using these oscillator
solutions~\cite{knp86}.  This allows to use this oscillator system
for wave functions in the Lorentz-covariant world.

However, presently, we are interested in space-time localizations of
the wave function dictated by the Gaussian factor or the ground-state
wave function.  In the light-cone coordinate system, the Lorentz-boosted
wave function becomes
\begin{equation}\label{eta2}
\psi_{\eta }(z,t) = \left({1 \over \pi }\right)^{1/2}
\exp\left\{-{1\over 2}\left(e^{-2\eta }u^{2} +
e^{2\eta}v^{2}\right)\right\} ,
\end{equation}
as given in Eq.(\ref{eta}).
This wave function can  be written as
\begin{equation}\label{wf03}
\psi_{\eta }(z,t) = \left({1 \over \pi }\right)^{1/2}
\exp\left\{-{1\over 4}\left[e^{-2\eta }(z + t)^{2} +
e^{2\eta}(z - t)^{2}\right]\right\} .
\end{equation}
Let us go back to Eq.(\ref{wf02}) for the coupled oscillators.
If we replace $x_{1}$ and $x_{2}$ by $z$ and $t$ respectively,
we arrive at the above expression for covariant harmonic
oscillators.

We of course talk about two different physical systems.  For the
case of coupled oscillators, there are two one-dimentional
oscillators.  In the case of covariant harmonic oscillators,
there is one oscillator with two variables.  The Lorentz boost
corresponds to coupling of two oscillators.  With these points
in mind, we can translate the physics of coupled oscillators into
the physics of covariant harmonic oscillators.

\section{Entangled Space and Time}\label{sten}

Let us now compare the space-time wave function of Eq.(\ref{wf03})
with the wave function Eq.(\ref{wf02}) for the coupled oscillators.
We can obtain the latter by replacing $x_{1}$ and $x_{2}$ in
the coupled-oscillator wave function by $z$ and $t$ respectively.

\begin{equation}\label{expan2}
\psi_{\eta }(z, t) = {1 \over \cosh\eta}\sum^{}_{k}
(\tanh\eta )^{k} \phi_{k}(z) \phi_{k}(t) ,
\end{equation}
This expansion is identical to that for the coupled oscillators if
$z$ and $t$ are replaced by $x_{1}$ and $x_{2}$ respectively.

Thus the space variable $z$ and the time variable $t$ are entangled
in the manner same as given in Ref.~\cite{giedke03}.  However, there
is a very important difference.  The $z$ variable is well defined in
the present form of quantum mechanics, but the time-separation
variable $t$ is not.  First of all, it is different from the calendar
time.  It exists because the simultaneity in special relativity is
not invariant in special relativity.  This point has not yet been
systematically examined.

All we can say at this point is that the Lorentz-entanglement requires
one variable we can measure, and the other variable we do not pretend
to measure. In his book on statistical mechanics~\cite{fey72}, Feynman
makes the following statement about the density matrix. {\it When we
solve a quantum-mechanical problem, what we really do is divide the
universe into two parts - the system in which we are interested and
the rest of the universe.  We then usually act as if the system in
which we are interested comprised the entire universe.  To motivate
the use of density matrices, let us see what happens when we include
the part of the universe outside the system}.

Does this time-separation variable exist when the hadron is at rest?
Yes, according to Einstein.  In the present form of quantum mechanics,
we pretend not to know anything about this variable.  Indeed, this
variable belongs to Feynman's rest of the universe.

We can use the coupled harmonic oscillators to illustrate what Feynman
says in his book.  Here we can use $x_{1}$ and $x_{2}$ for the variable
we observe and the variable in the rest of the universe.  By using the
rest of the universe, Feynman does not rule out the possibility of other
creatures measuring the $x_{2}$ variable in their part of the universe.

Using the wave function $\psi_{\eta}(z,t)$ of Eq.(\ref{wf02}),
we can construct the pure-state density matrix
\begin{equation}
\rho(z,t;z',t')
= \psi_{\eta}(z,t)\psi_{\eta}(z',t') ,
\end{equation}
which satisfies the condition $\rho^{2} = \rho $:
\begin{equation}
\rho(z,t;z',t') =
\int \rho(z,t;z'',t'')
\rho(z'',t'';z',t') dz'' dt'' .
\end{equation}
If we are not able to make observations on $t$, we should
take the trace of the $\rho$ matrix with respect to the $t$
variable.  Then the resulting density matrix is
\begin{equation}\label{integ}
\rho(z, z') = \int \rho (z,t;x'_{1},t) dt .
\end{equation}

The above density matrix can also be calculated from the expansion of
the wave function given in Eq.(\ref {expan1}).  If we perform the integral
of Eq.(\ref{integ}), the result is
\begin{equation}\label{dmat}
\rho(z,z') = \left({1 \over \cosh(\eta)}\right)^{2}
\sum^{}_{k} (\tanh\eta)^{2k}
\phi_{k}(z)\phi^{*}_{k}(z') .
\end{equation}
The trace of this density matrix is $1$.  It is also straightforward to
compute the integral for $Tr(\rho^{2})$.  The calculation leads to
\begin{equation}
Tr\left(\rho^{2} \right)
= \left({1 \over \cosh(\eta)}\right)^{4}
\sum^{}_{k} (\tanh\eta)^{4k} .
\end{equation}
The sum of this series is $1/\cosh(2\eta)$ which is less than one.

This is of course due to the fact that we are averaging over the $x_{2}$
variable which we do not measure.  The standard way to measure this
ignorance is to calculate the entropy defined as
\begin{equation}
S = - Tr\left(\rho \ln(\rho) \right) ,
\end{equation}
where $S$ is measured in units of Boltzmann's constant.  If we use the
density matrix given in Eq.(\ref{dmat}), the entropy becomes
\begin{equation}
S = 2 \left\{\cosh^{2}\eta \ln(\cosh\eta) -
             \sinh^{2}\eta \ln(\sinh\eta)    \right\} .
\end{equation}
This expression can be translated into a more familiar form if
we use the notation
\begin{equation}
\tanh\eta  = \exp\left(-{\hbar\omega \over kT}\right) ,
\end{equation}
where $\omega$ is the unit of energy spacing, and $k$ and $T$ are
Boltzmann's constant and absolute Temperature respectively.  The
ratio $\hbar\omega/kT$ is a dimensionless variable.  In terms of
this variable, the entropy takes the form~\cite{hkn89pl}
\begin{equation}
S = \left({\hbar\omega \over kT}\right)
\frac{1}{\exp(\hbar\omega/kT) - 1}
- \ln\left[1 - \exp(-\hbar\omega/kT)\right] .
\end{equation}
This familiar expression is for the entropy of an oscillator state
in thermal equilibrium.  Thus, for this oscillator system, we can
relate our ignorance of the time-separation variable to the
temperature.  It is interesting to note that the boost parameter or
coupling strength measured by $\eta$ can be related to a temperature
variable.

Let us summarize.  At this time, the only theoretical tool available
to this time-separation variable is through the space-time
entanglement, which generate entropy coming from the rest of the
universe.  If the time-separation variable is not measured the
entropy is one of the variables to be taken into account in the
Lorentz-covariant system.

In spite of our ignorance about this time-separation variable
from the theoretical point of view, its existence has been proved
beyond any doubt in high-energy laboratories.  We shall see in
Sec.~\ref{feydeco}.that it plays a role in producing a decoherence
effect observed universally in high-energy laboratories.

\section{Feynman's Decoherence}\label{feydeco}

In a hydrogen atom or a hadron consisting of two quarks, there is a
spacial separation between two constituent elements.  In the case of
the hydrogen we call it the Bohr radius.  It the atom or hadron is
at rest, the time-separation variable does not play any visible role
in quantum mechanics.  However, if the system is boosted to the
Lorentz frame which moves with a speed close to that of light, this
time-separation variable becomes as important as the space separation
of the Bohr radius.  Thus, the time-separation plays a visible role
in high-energy physics which studies fast-moving bound states.  Let
us study this problem in more detail.

It is a widely accepted view that hadrons are quantum bound states
of quarks having localized probability distribution.  As in all
bound-state cases, this localization condition is responsible for
the existence of discrete mass spectra.  The most convincing evidence
for this bound-state picture is the hadronic mass
spectra~\cite{fkr71,knp86}.
However, this picture of bound states is applicable only to observers
in the Lorentz frame in which the hadron is at rest.  How would the
hadrons appear to observers in other Lorentz frames?

In 1969, Feynman observed that a fast-moving hadron can be regarded
as a collection of many ``partons'' whose properties do not appear
to be quite different from those of the quarks~\cite{fey69}.  For
example, the number of quarks inside a static proton is three, while
the number of partons in a rapidly moving proton appears to be infinite.
The question then is how the proton looking like a bound state of
quarks to one observer can appear different to an observer in a
different Lorentz frame?  Feynman made the following systematic
observations.

\begin{itemize}

\item[a.]  The picture is valid only for hadrons moving with
  velocity close to that of light.

\item[b.]  The interaction time between the quarks becomes dilated,
   and partons behave as free independent particles.

\item[c.]  The momentum distribution of partons becomes widespread as
   the hadron moves fast.

\item[d.]  The number of partons seems to be infinite or much larger
    than that of quarks.

\end{itemize}

\noindent Because the hadron is believed to be a bound state of two
or three quarks, each of the above phenomena appears as a paradox,
particularly b) and c) together.  How can a free particle have a
wide-spread momentum distribution?

\begin{figure}
\centerline{\includegraphics[scale=0.5]{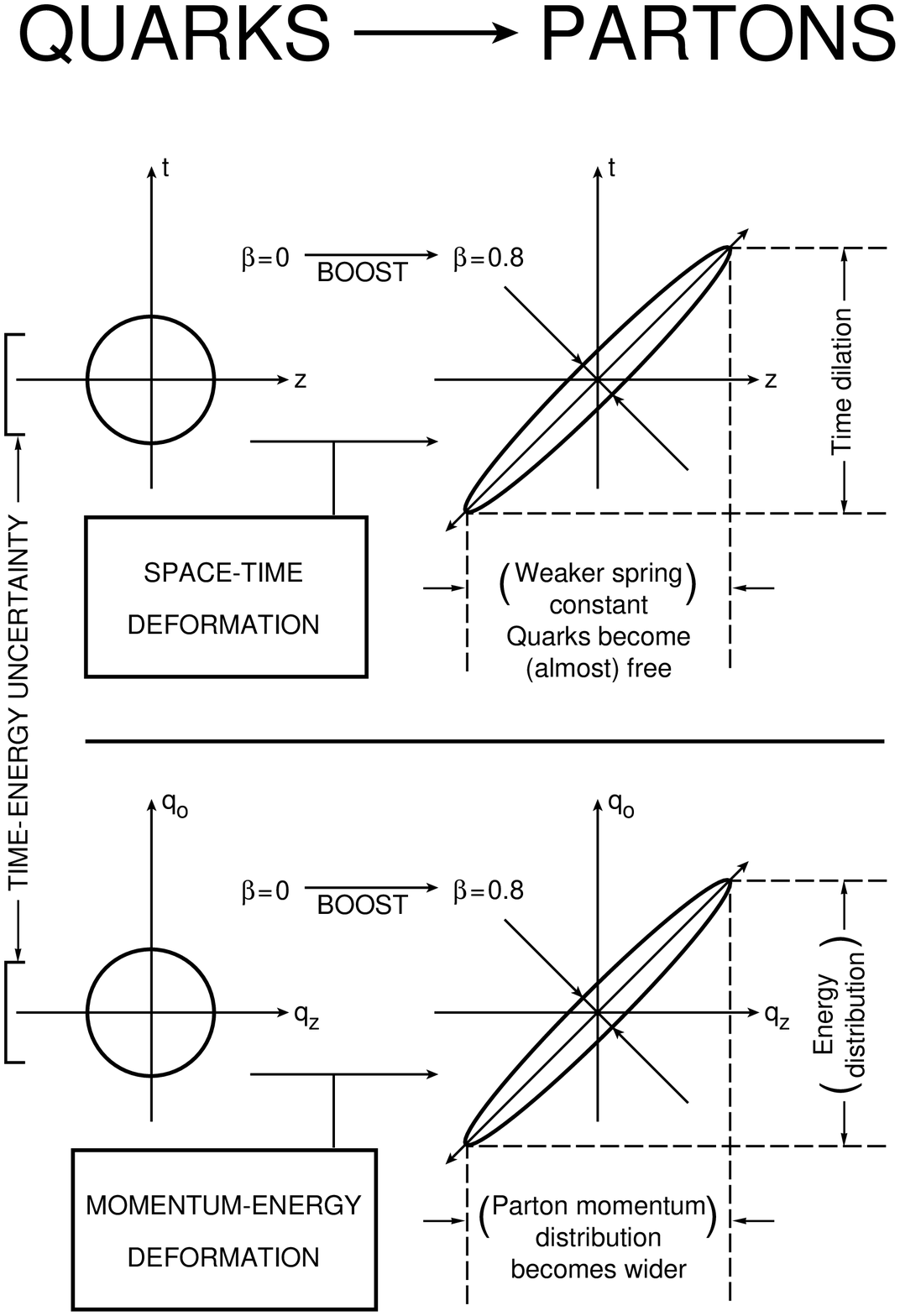}}
\vspace{5mm}
\caption{Lorentz-squeezed space-time and momentum-energy wave
functions.  As the hadron's speed approaches that of light, both
wave functions become concentrated along their respective positive
light-cone axes.  These light-cone concentrations lead to Feynman's
parton picture.}\label{parton}
\end{figure}

In order to resolve this paradox, let us construct the
momentum-energy wave function corresponding to Eq.(\ref{eta}).
If the quarks have the four-momenta $p_{a}$ and $p_{b}$, we can
construct two independent four-momentum variables~\cite{fkr71}
\begin{equation}
P = p_{a} + p_{b} , \qquad q = \sqrt{2}(p_{a} - p_{b}) .
\end{equation}
The four-momentum $P$ is the total four-momentum and is thus the
hadronic four-momentum.  $q$ measures the four-momentum separation
between the quarks.  Their light-cone variables are
\begin{equation}\label{conju}
q_{u} = (q_{0} - q_{z})/\sqrt{2} ,  \qquad
q_{v} = (q_{0} + q_{z})/\sqrt{2} .
\end{equation}
The resulting momentum-energy wave function is
\begin{equation}\label{phi}
\phi_{\eta }(q_{z},q_{0}) = \left({1 \over \pi }\right)^{1/2}
\exp\left\{-{1\over 2}\left[e^{-2\eta}q_{u}^{2} +
e^{2\eta}q_{v}^{2}\right]\right\} .
\end{equation}
Because we are using here the harmonic oscillator, the mathematical
form of the above momentum-energy wave function is identical to that
of the space-time wave function of Eq.(\ref{eta}).  The Lorentz
squeeze properties of these wave functions are also the same.  This
aspect of the squeeze has been exhaustively discussed in the
literature~\cite{knp86,kn77par,kim89}, and they are illustrated again
in Fig.~\ref{parton} of the present pape.  The hadronic structure
function calculated from this formalism is in a reasonable agreement
with the experimental data~\cite{hussar81}.

When the hadron is at rest with $\eta = 0$, both wave functions
behave like those for the static bound state of quarks.  As $\eta$
increases, the wave functions become continuously squeezed until
they become concentrated along their respective positive
light-cone axes.  Let us look at the z-axis projection of the
space-time wave function.  Indeed, the width of the quark distribution
increases as the hadronic speed approaches that of the speed of
light.  The position of each quark appears widespread to the observer
in the laboratory frame, and the quarks appear like free particles.

The momentum-energy wave function is just like the space-time wave
function.  The longitudinal momentum distribution becomes wide-spread
as the hadronic speed approaches the velocity of light.  This is in
contradiction with our expectation from nonrelativistic quantum
mechanics that the width of the momentum distribution is inversely
proportional to that of the position wave function.  Our expectation
is that if the quarks are free, they must have their sharply defined
momenta, not a wide-spread distribution.

However, according to our Lorentz-squeezed space-time and
momentum-energy wave functions, the space-time width and the
momentum-energy width increase in the same direction as the hadron
is boosted.  This is of course an effect of Lorentz covariance.
This indeed is to the resolution of one of the the quark-parton
puzzles~\cite{knp86,kn77par,kim89}.

Another puzzling problem in the parton picture is that partons appear
as incoherent particles, while quarks are coherent when the hadron
is at rest.  Does this mean that the coherence is destroyed by the
Lorentz boost?   The answer is NO, and here is the resolution to
this puzzle.

When the hadron is boosted, the hadronic matter becomes squeezed and
becomes concentrated in the elliptic region along the positive
light-cone axis.  The length of the major axis becomes expanded by
$e^{\eta}$, and the minor axis is contracted by $e^{\eta}$.

This means that the interaction time of the quarks among themselves
become dilated.  Because the wave function becomes wide-spread, the
distance between one end of the harmonic oscillator well and the
other end increases.  This effect, first noted by Feynman~\cite{fey69},
is universally observed in high-energy hadronic experiments.  The
period of oscillation is increases like $e^{\eta}$.

\begin{figure}

\centerline{\includegraphics[scale=0.4]{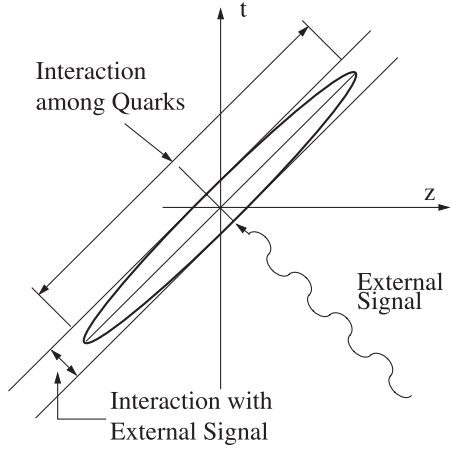}}
\caption{Quarks interact among themselves and with external signal.
The interaction time of the quarks among themselves become dilated,
as the major axis of this ellipse indicates.  On the other hand, the
the external signal, since it is moving in the direction opposite to
the direction of the hadron, travels along the negative light-cone
axis.  To the external signal, if it moves with velocity of light,
the hadron appears very thin, and the quark's interaction time with
the external signal becomes very small.}\label{tdil}

\end{figure}

On the other hand, the external signal, since it is moving in the
direction opposite to the direction of the hadron travels along
the negative light-cone axis, as illustrated in Fig.~\ref{tdil}.

If the hadron contracts along the negative light-cone axis, the
interaction time decreases by $e^{-\eta}$.  The ratio of the interaction
time to the oscillator period becomes $e^{-2\eta}$.  The energy of each
proton coming out of the Fermilab accelerator is $900 GeV$.  This leads
the ratio to $10^{-6}$.  This is indeed a small number.  The external
signal is not able to sense the interaction of the quarks among
themselves inside the hadron.

Indeed, Feynman's parton picture is one concrete physical example
where the decoherence effect is observed.  As for the entropy, the
time-separation variable belongs to the rest of the universe.  Because
we are not able to observe this variable, the entropy increases
as the hadron is boosted to exhibit the parton effect.  The
decoherence is thus accompanied by an entropy increase.

Let us go back to the coupled-oscillator system.  The light-cone
variables in Eq.(\ref{eta}) correspond to the normal coordinates in
the coupled-oscillator system given in Eq.(\ref{normal}).  According
to Feynman's parton picture, the decoherence mechanism is determined
by the ratio of widths of the wave function along the two normal
coordinates.


This decoherence mechanism observed in Feynman's parton picture is quite
different from other dicoherences discussed in the literature.  It is
widely understood that the word decoherence is the loss of coherence within
a system.  On the other hand, Feynman's decoherence discussed in this
section comes from the way external signal interacts with the internal
constituents.

\section*{Concluding Remarks}
In this paper, we noted first that two-mode squeezed states can play a major
role in clarifying some of the entanglement ideas.  Since the mathematical
language of two-mode states is that of two coupled oscillators, the
oscillator system can be a reservoir of physical ideas associated with
entanglements.  Then, other physical models derivable from the coupled
oscillators can carry the physics of entanglement.

We have shown in this paper, the covariant harmonic oscillator system
with one space and one time variable share the same mathematical
framework as the coupled harmonic oscillators.  Thus, the oscillator
system gives a concrete example of space-time entanglement.

Thanks to its Lorentz covariance, the covariant oscillator system can
explain the quark model and parton model as two limiting cases of the
same covariant entity.  It can explain the peculiarities observed in
Feynman's parton picture.  The most controversial aspect in the
parton model is that, while the quarks interact coherently with
external signals, partons behave like free particles interacting
without coherence with external signals.  This phenomenon was
observed first by Feynman.  Thus, it is quite appropriate to call this
Feynman's decoherence.  In this paper, we have provided a resolution
to this parton puzzle.  It requires a space-time picture of entanglement.

\section*{Acknowledgments}
We would like to thank G. S. Agarwal, H. Hammer, and A. Vourdas for
helpful discussion on the precise definition of the word ``entanglement''
applicable to coupled systems.


\end{document}